\documentclass[a4paper,aps,prd,10pt,preprintnumbers,twocolumn,groupedaddress,nofootinbib,amsmath,amssymb,showkeys]{revtex4-2}
\usepackage{ragged2e}
\usepackage{graphicx}
\usepackage[utf8]{inputenc}
\usepackage[T1]{fontenc}
\usepackage{cmap}
\usepackage{float}
\usepackage{amsmath}

\DeclareMathAlphabet{\pazocal}{OMS}{zplm}{m}{n}

\usepackage[colorlinks=true, linkcolor=blue, citecolor=blue, urlcolor=blue]{hyperref}
\begin{document}
\title{Photon regions, shadow observables and constraints from M87* of a Kerr-Newman-like black hole in Bumblebee gravity surrounded by plasma}
\author{Jian-Peng Zhang}

\author{Yu Zhang} \email{zhangyu_128@126.com (Corresponding author)}

\author{Li Han}
\affiliation{Faculty of Science, Kunming University of Science and Technology, Kunming, Yunnan 650500, China}

\begin{abstract}
In this paper, we investigate the photon regions, shadow, and observational constraints of a Kerr-Newman-like black hole in Bumblebee gravity within a plasma medium. By employing a specific non-homogeneous power-law plasma model to ensure the separability of the Hamilton-Jacobi equation, we derive the null geodesic equations, analyze the photon regions, and construct the black hole shadow. Furthermore, we introduce two sets of shadow observables to systematically analyze the distinct effects of each physical parameter (spin $a$, charge $Q_0$, Lorentz-violating parameter $\ell$, and plasma parameter $k$) on the shadow geometry. Specifically, we find that $a$ and $\ell$ mainly enhance the distortion of the shadow, whereas $Q_0$ and $k$ primarily lead to its radial shrinkage.  Additionally, a brief evaluation of the energy emission rate shows that an increase in these parameters generally suppresses the emission peak. Finally, by modeling M87* as a charged rotating black hole in Bumblebee gravity surrounded by plasma, we can constrain the physical parameters using observations from the Event Horizon Telescope (EHT). While the angular diameter $\theta_d = 42 \pm 3 \, \mu\text{as}$ narrows the viable parameter space, the circularity deviation $\Delta C \lesssim 0.1$ and axis ratio $1 < D_x \lesssim 4/3$ obey the EHT limits. This suggests that the charged rotating black hole in Bumblebee gravity surrounded by plasma might be a candidate for real astrophysical black holes.

\vspace{1em}
\end{abstract}
\maketitle

\section{Introduction}
\label{sec:intro}
The successful capture of the M87* \cite{EventHorizonTelescope:2019dse,EventHorizonTelescope:2019uob,EventHorizonTelescope:2019jan,EventHorizonTelescope:2019ths,EventHorizonTelescope:2019pgp,EventHorizonTelescope:2019ggy} and Sgr A* \cite{EventHorizonTelescope:2022wkp,EventHorizonTelescope:2022apq,EventHorizonTelescope:2022wok} images by the Event Horizon Telescope (EHT) collaboration marks a new era in which black hole astrophysics enters the stage of testing strong gravity theories through direct imaging \cite{Psaltis:2018xkc,Glampedakis:2021oie,Vincent:2020dij}. 
The black hole shadow, characterized by a central dark region surrounded by a bright ring, is fundamentally the optical projection of the spacetime geometry near the black hole onto the celestial sphere of a distant observer. Although current EHT observations are highly consistent with the predictions of the Kerr black hole in General Relativity (GR) within error margins \cite{EventHorizonTelescope:2019pgp}, they still leave room for exploring new physics beyond the GR within the allowable observational precision, such as dark matter \cite{Hui:2021tkt,EPTA:2023xxk,NANOGrav:2023hvm}, modified gravity theories \cite{Clifton:2011jh,Fernandes:2022zrq,Shankaranarayanan:2022wbx}, and quantum gravity effects \cite{Han:2020uhb,Zhang:2023okw,Konoplya:2019xmn}.

Against this backdrop, using black hole shadows to test potential deviations from GR has become one of the most important avenues for studying modified gravity theories \cite{Uniyal:2022vdu,Silva:2022srr,Atamurotov:2022knb,Ayzenberg:2018jip,Wang:2025fmz,Kuang:2022ojj}. Among the various candidate theories, those involving Lorentz symmetry breaking (LSB) have attracted widespread attention due to their clear physical motivations \cite{Kostelecky:2003fs,Ding:2021iwv,Karmakar:2023mhs}. Although Lorentz symmetry serves as a fundamental cornerstone of both quantum field theory and GR, its strict validity at all energy scales remains an open question. Various candidate theories of quantum gravity predict that this symmetry might be broken at high-energy scales \cite{Kostelecky:1988zi,Jacobson:2005bg}, leaving observable effects in low-energy effective theories \cite{Kostelecky:2016kfm}. To systematically characterize such LSB effects, Kostelecky et al. \cite{Colladay:1998fq} developed the Standard Model Extension framework, providing a unified framework for describing Lorentz violation. In the gravity sector, the Einstein-Bumblebee model stands as one of the simplest implementations of the Standard Model Extension \cite{Kostelecky:1988zi,Kostelecky:1989jp,Kostelecky:2010ze}. It induces spontaneous LSB by introducing a vector field with a non-zero vacuum expectation value. In recent years, Schwarzschild-like \cite{Casana:2017jkc,Gullu:2020qzu,Bailey:2025oun} and Kerr-like black hole solutions \cite{Ding:2019mal,Ding:2020kfr,Jha:2020pvk} have been derived within this framework and utilized to investigate horizon structures and photon geodesic properties. However, most existing studies on the shadow of bumblebee black holes are confined to the vacuum conditions \cite{Uniyal:2022xnq,Kuang:2022xjp,Pantig:2024lpg,Islam:2024sph,Gao:2024ejs,Sekhmani:2025gvv}, neglecting the significant effects potentially imposed by realistic astrophysical environments.

On the other hand, even without invoking corrections to GR, the interpretation of shadow observations is strictly constrained by environmental factors. In realistic astrophysical scenarios, black holes are surrounded by accretion flows, jets, or the interstellar medium \cite{Yuan:2014gma,Perlick:2015vta}, which on macroscopic scales can be effectively modeled as a plasma medium. As noted in previous studies \cite{Bisnovatyi-Kogan:2010flt,Sareny:2019kxs}, the dispersive properties of plasma alter the dispersion relation of photons, forcing their trajectories to deviate from standard vacuum null geodesics. Following this, the pioneering work of Atamurotov et al. \cite{Atamurotov:2015nra} systematically formulated the theoretical framework describing plasma effects on black hole shadows. Subsequent research \cite{Abdujabbarov:2016hnw,Bisnovatyi-Kogan:2017kii,Perlick:2017fio,Zhang:2022osx,Chowdhuri:2020ipb,Li:2021btf} has demonstrated that, whether the plasma distribution is homogeneous or non-homogeneous, the primary consequence is a systematic shrinkage of the shadow's apparent size, accompanied by deformations in its shape.

Current research on this problem has primarily proceeded along two paths: one investigates the impact of modified gravity theories on black hole shadows, yet is often confined to the vacuum conditions \cite{Sun:2024xtf,Nozari:2024jiz,Meng:2022kjs,Heydari-Fard:2023ent,Liu:2024lve}, while the other analyzes plasma correction effects, but strictly within the framework of GR \cite{Babar:2020txt,Badia:2021kpk,Atamurotov:2022iwj,Molla:2022izk,Pahlavon:2024caj}. Consequently, for scenarios incorporating both modified gravity effects and realistic plasma environments \cite{Badia:2022phg,Ali:2023eqj,Wang:2021irh}, systematic analyses of black hole shadow and its observables within a unified theoretical framework remain relatively limited.

In this paper, we investigate a Kerr-Newman-like (KN-like) black hole in Bumblebee gravity, immersed in a non-homogeneous plasma background. To ensure computational feasibility while maintaining physical consistency, we adopt a radial power-law non-homogeneous plasma model that preserves the separability of the Hamilton-Jacobi equation. The horizons, null geodesic equations, and photon region of the black hole are shown. Within this framework, we systematically calculate the shadow and introduce two sets of shadow observables to characterize the influence of various physical parameters. Furthermore, utilizing the EHT observations of M87* --- specifically the angular diameter, axial ratio, and circularity deviation --- we place stringent constraints on the parameter space of this coupled model. This enables us to test the viability of the KN-like black hole as an M87* candidate within a realistic plasma environment.

This paper is structured as follows: In section~\ref{Black hole within the framework of Bumblebee gravity}, we introduce the metric of the KN-like black hole, discussing its horizons and the valid parameter range. Section~\ref{Null geodesics and photon regions in a plasma environment} derives of the null geodesic equations in the presence of plasma and subsequently analyzes the corresponding photon regions to determine the boundary of the shadow. Section~\ref{The black hole shadow and shadow observables} presents the black hole shadows and characterizes how the shadow shape and observables respond to parameter variations, while also providing a brief evaluation of the associated energy emission rate. In section~\ref{Constraints from EHT Observations on M87*}, we impose constraints on the model parameters by confronting our theoretical results with the EHT observational data. Finally, a summary is provided in section~\ref{Conclusion}. Throughout this paper, unless otherwise specified, we employ geometric units where $G=c=M=1$ and adopt the metric signature (-, +, +, +).

\section{Black hole within the framework of Bumblebee gravity}
\label{Black hole within the framework of Bumblebee gravity}

\begin{figure*}[t]
\centering
\includegraphics[width=0.45\textwidth]{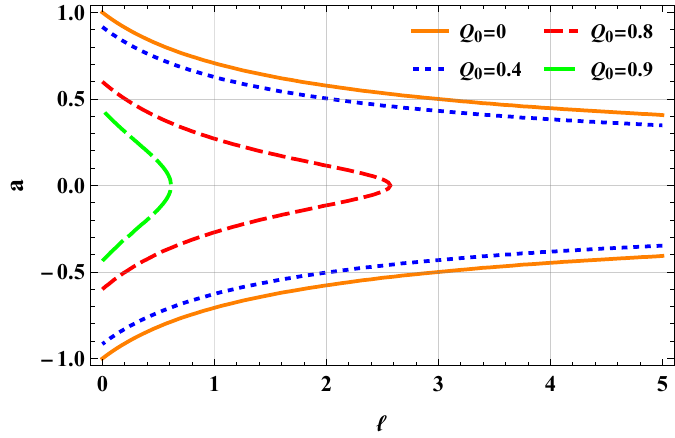}
\includegraphics[width=0.45\textwidth]{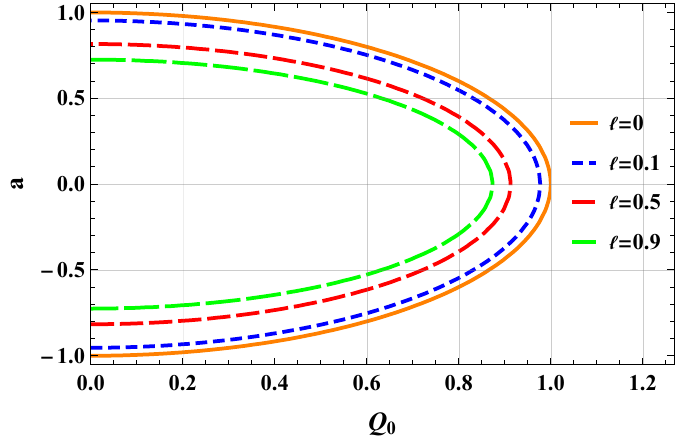}
\caption{Parameter space of the Bumblebee black hole ($M=1$). The boundary of the curve represents represent the extremal black hole condition, corresponding to the degenerate roots of the radial metric function. The region enclosed by the curves denotes the parameter space where an event horizon exists, whereas the exterior region corresponds to naked singularities.}
\label{fig1}
\end{figure*}

The black hole solution considered in this paper is based on the Bumblebee gravity theory. In this theory \cite{Kostelecky:1988zi,Kostelecky:1989jp,Kostelecky:2010ze}, a vector field $B_u$ with a non-zero vacuum expectation value is introduced to induce spontaneous Lorentz symmetry breaking (LSB) in the gravitational interaction. Consequently, a parameter characterizing the breaking strength appears in the low-energy effective field theory.

\begin{figure*}[t]
\centering
\includegraphics[width=0.32\textwidth]{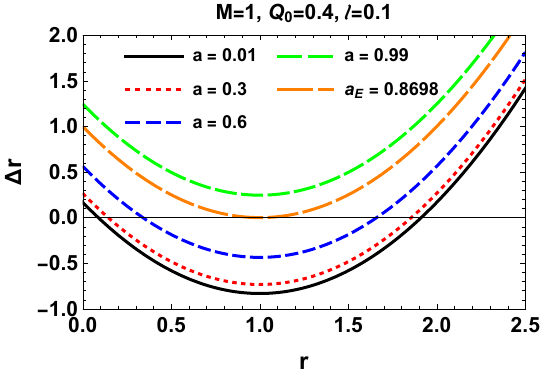}
\includegraphics[width=0.32\textwidth]{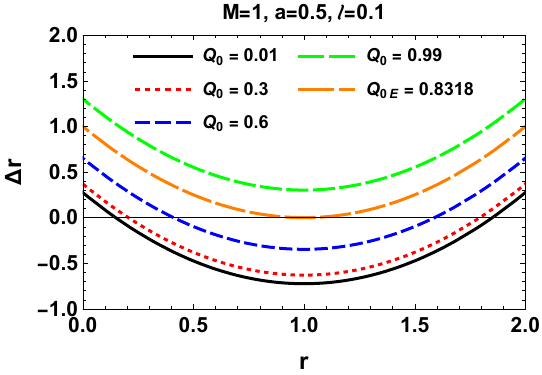}
\includegraphics[width=0.32\textwidth]{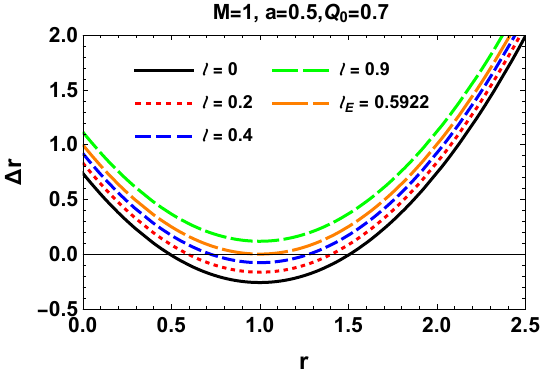}

\includegraphics[width=0.32\textwidth]{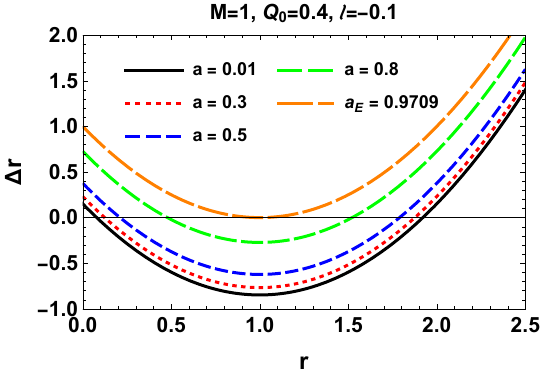}
\includegraphics[width=0.32\textwidth]{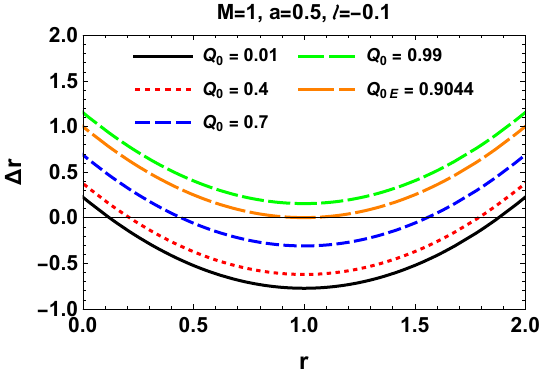}
\includegraphics[width=0.32\textwidth]{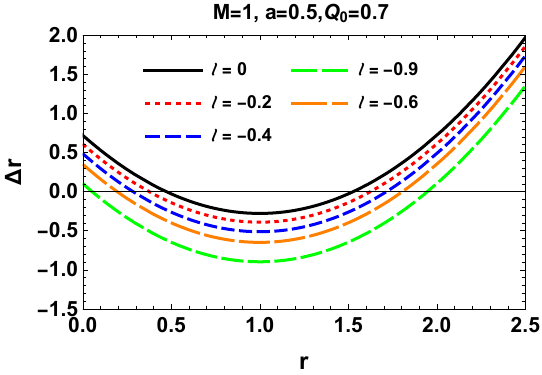}
\caption{Plots of the radial function $\Delta_r(r)$ versus $r$ for different parameter combinations ($M=1$). The columns (from left to right) display the effects of spin $a$, charge $Q_0$, and Lorentz-violation parameter $\ell$, while the upper and lower rows correspond to positive and negative $\ell$, respectively. The roots of $\Delta_r(r)=0$ denote the Cauchy (left) and event (right) horizons, with tangency indicating extremal black holes.}
\label{fig2}
\end{figure*}

In this work, we adopt the stationary, axisymmetric, and charged Kerr-Newman-like (KN-like) black hole solution recently obtained by Liu et al. \cite{Liu:2024axg}. Expressed in Boyer-Lindquist coordinates, the metric form is given by
\begin{equation}\label{2.1}
\begin{aligned}
ds^{2} ={}&
-\frac{\Delta_{r}}{\rho^{2}}
\left(dt-a\sqrt{1+\ell}\sin^{2}\theta\,d\phi\right)^{2} \\
&+(1+\ell)\frac{\rho^{2}}{\Delta_{r}}\,dr^{2}
+\rho^{2}d\theta^{2} \\
&+\frac{\sin^{2}\theta}{\rho^{2}}
\left(a\sqrt{1+\ell}\,dt-\left(r^{2}+a^{2}(1+\ell)\right)d\phi\right)^{2},
\end{aligned}
\end{equation}
where the relevant metric functions are defined as

\begin{equation}\label{2.2}
\begin{split}
\Delta_{r} &= (1+\ell)a^{2}+r^{2}-2Mr+\frac{2(1+\ell)Q_{0}^{2}}{2+\ell}, \\
\rho^{2} &= r^{2}+(1+\ell)a^{2}\cos^{2}\theta.
\end{split}
\end{equation}

Here, $M$ denotes the black hole mass, $a$ represents the spin parameter, and $Q_{0}$ is the charge parameter. The Lorentz-violating parameter $\ell$ characterizes the strength of the LSB. When $\ell \to 0$, the metric reduces to the Kerr-Newman solution in general relativity. If we further set $Q_{0}\to 0$, it recovers the standard Kerr black hole solution.

The horizons of a black hole are determined by the roots of its radial metric function, which satisfy
\begin{equation}\label{2.4}
\Delta_{r}=(1+\ell)a^{2}+r^{2}-2Mr+\frac{2(1+\ell)Q_{0}^{2}}{2+\ell}=0.
\end{equation}

This equation corresponds to three root structures: two positive real roots, one double root and no positive real roots. Specifically, the existence of two distinct positive real roots corresponds to a black hole with an event horizon and a Cauchy horizon. When the two roots coincide, it corresponds to an extremal black hole. Finally, if no positive real roots exist, the horizon vanishes, and the spacetime evolves into a naked singularity. Figure~\ref{fig1} presents the boundaries of the extremal black hole in the $(\ell, a)$ and $(Q_0, a)$ parameter planes. It is obvious that as the charge parameter $Q_0$ or the Lorentz-violation parameter $\ell$ increases, the allowable range for the spin parameter $a$ shrinks. 

Figure~\ref{fig2} shows the behavior of the radial metric function $\Delta_r(r)$ as a function of the radial coordinate for different parameter combinations. The intersection of the curves with the $r$-axis corresponds to the horizon positions, where the left and right intersections represent the Cauchy horizon and the event horizon, respectively, while the tangent point corresponds to the extremal black hole case. It can be seen that increasing the spin parameter $a$, the charge parameter $Q_0$, and the Lorentz-violating parameter $\ell$ cause the Cauchy horizon and the event horizon to approach each other. Based on the above analysis, all subsequent calculations of the black hole shadow are restricted to the parameter region that ensures the existence of the event horizon.

\section{Null geodesics and photon regions in a plasma environment}
\label{Null geodesics and photon regions in a plasma environment}
To investigate the joint effects of Lorentz symmetry breaking (LSB) in Bumblebee gravity and the plasma environment on the shadow structure, it is necessary to characterize the photon propagation behavior and photon regions in this background. Since the spacetime under consideration is stationary and axisymmetric, the photon motion problem can be formulated as a Hamiltonian dynamical system with a structure of conserved quantities. In the study of shadows from axisymmetric black holes, the Hamilton-Jacobi (H-J) equation is widely used due to its advantage of separability of variables.

However, the introduction of a plasma medium causes the photon dispersion relation to significantly deviate from the vacuum case, generally disrupting the separability of the H-J equation. Therefore, selecting a plasma model that has a clear astrophysical motivation while maintaining the separability of the H-J equation is crucial. This paper employs a specific non-homogeneous power-law plasma model, derives the null geodesic equations of photons on this basis, and analyzes the photon regions, laying the foundation for the subsequent construction of black hole shadows.

In the vacuum background, the motion of photons is described by the H-J equation
\begin{equation}\label{3.1}
H(x, p) = \frac{1}{2} g^{\mu\nu}(x) p_\mu p_\nu = 0, 
\end{equation}
where $g^{\mu\nu}$ is the inverse metric and $p_\mu$ is the photon's covariant four-momentum.

When considering a pressureless, non-magnetized plasma, the Hamiltonian for photons needs to include a plasma frequency term, which takes the form:
\begin{equation}\label{3.2}
H(x, p) = \frac{1}{2} (g^{\mu\nu}(x) p_\mu p_\nu + \omega_p(x)^2) = 0, 
\end{equation}
where $\omega_p$ denotes the plasma electron frequency.

Generally, an arbitrary spatial distribution of the plasma frequency tends to break the separability of the H-J equation. Previous studies \cite{Badia:2021kpk,Bezdekova:2022gib,Fathi:2021mjc} have indicated that the variable separation satisfies the condition:
\begin{equation}\label{3.3}
\omega_p^2(r, \theta) = \frac{f_r(r) + f_\theta(\theta)}{r^2 + (1+\ell)a^2 \cos^2\theta}, 
\end{equation}
where $f_r(r)$ and $f_\theta(\theta)$ are functions of their respective coordinates. This condition, first proposed by Perlick and Tsupko \cite{Perlick:2017fio} in Kerr spacetime, has been widely adopted to investigate the shadow of axisymmetric black holes in plasma environments.

Given this separable structure and the stationarity and axisymmetry of the spacetime, the photon energy $E$ and axial angular momentum $L_z$ are conserved quantities, defined as: $E = -p_t$, $L_z = p_\phi$. Accordingly, we adopt the standard separable ansatz for the Action $S$:
\begin{equation}\label{3.4}
S = -Et + L_z\phi + S_r(r) + S_\theta(\theta), 
\end{equation}
where $S_r$ and $S_\theta$ represent the radial and angular parts of the action, respectively. By substituting this expression into eq.\eqref{3.2}, we successfully separate the photon motion into independent radial and angular equations:
\begin{equation}\label{3.5}
(\rho^2 \frac{\partial S_r}{\partial r})^2 = R(r), \quad (\rho^2 \frac{\partial S_\theta}{\partial \theta})^2 = \Theta(\theta). 
\end{equation}

The resulting separation constant is identified as the generalized Carter constant $\mathcal{K}$:
\begin{equation}\label{3.6}
\begin{aligned}
\mathcal{K} &:= -\frac{\Delta_r}{1+\ell}\left(\rho^2 \frac{\partial S_r}{\partial r}\right)^2 \\
&\quad + \frac{[(r^2+a_{\text{eff}}^2)E - a_{\text{eff}}L_z]^2}{\Delta_r} - (L_z - a_{\text{eff}}E)^2 - f_r(r) \\
&= \left(\rho^2 \frac{\partial S_\theta}{\partial \theta}\right)^2 - \left[a_{\text{eff}}^2 E^2 - \frac{L_z^2}{\sin^2\theta}\right] \cos^2\theta + f_\theta(\theta).
\end{aligned}
\end{equation}

To simplify the subsequent discussion, we introduce the effective spin parameter ${a_{\mathrm{eff}} = \sqrt{1+\ell}}$.

The radial potential function $R(r)$ and angular potential function $\Theta(\theta)$ of a photon are defined as follows:
\begin{equation}\label{3.7}
\begin{aligned}
\frac{\Delta_r}{1+\ell} \rho^2 \dot{r}^2 &= \frac{\Delta_r}{1+\ell} R(r) \\
&= \frac{\left[ (r^2 + a_{\text{eff}}^2)E - a_{\text{eff}}L_z \right]^2}{\Delta_r} \\
&\quad - (L_z - a_{\text{eff}}E)^2 - \mathcal{K} - f_r(r), \\[1ex]
\rho^2 \dot{\theta}^2 &= \Theta(\theta) \\
&= \mathcal{K} + \left[ a_{\text{eff}}^2 E^2 - \frac{L_z^2}{\sin^2 \theta} \right] \cos^2 \theta - f_\theta(\theta).
\end{aligned}
\end{equation}

Shapiro's research \cite{Shapiro:1974} on the accretion environment of Kerr black holes indicates that, over a considerable radial range, the plasma density and associated frequency follow a power-law distribution: ${\rho \propto \omega_p^2 \propto r^{-3/2}}$.

Combining this astrophysical motivation with the separability condition eq.\eqref{3.3}, the following plasma model \cite{Badia:2021kpk} is selected:
\begin{equation}\label{3.8}
f_r(r) = \omega_c^2 \sqrt{r}, \quad f_\theta(\theta) = 0, 
\end{equation}
where $\omega_c$ is a constant. Inserting this model into the radial potential function eq.\eqref{3.7} leads to a dimensionless coupling factor:
\begin{equation}\label{3.9}
k \equiv \frac{\omega_c^2}{\omega_0^2}. 
\end{equation}

Here $\omega_0$ signifies the photon frequency measured by an observer at infinity, and $k$ serves as a parameter characterizing the plasma intensity.

\begin{figure*}[t]
\centering
\includegraphics[width=1\textwidth]{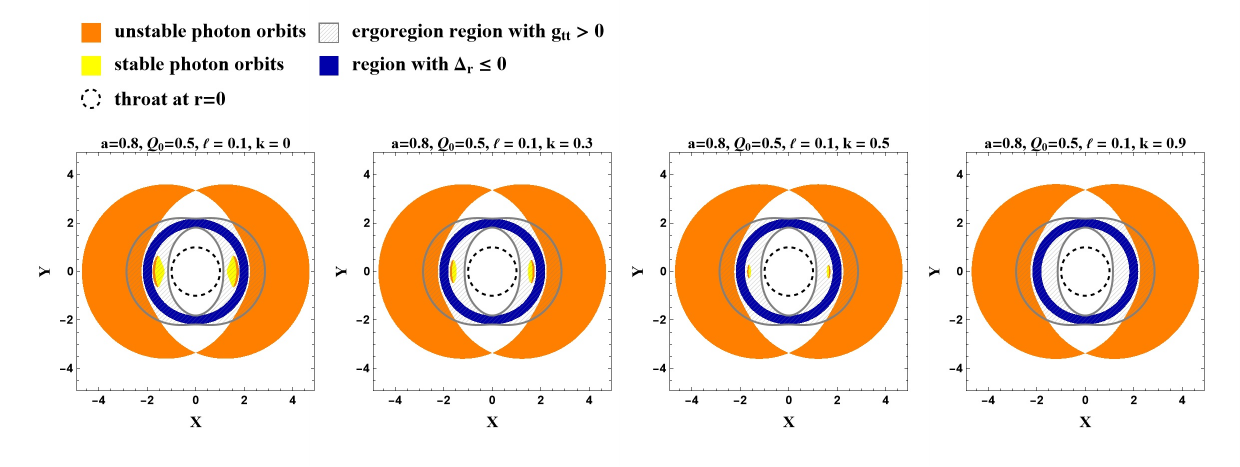}
\caption{The photon regions of the KN-like black hole in Bumblebee gravity surrounded by plasma in the $(r, \theta)$ plane. The parameters are set to $a = 0.8$, $Q_0 = 0.5$, and $\ell = 0.1$, with varying plasma parameter $k$.}
\label{fig3a}
\end{figure*}

\vspace{0.5 cm}

\begin{figure*}[t]
\centering
\includegraphics[width=1\textwidth]{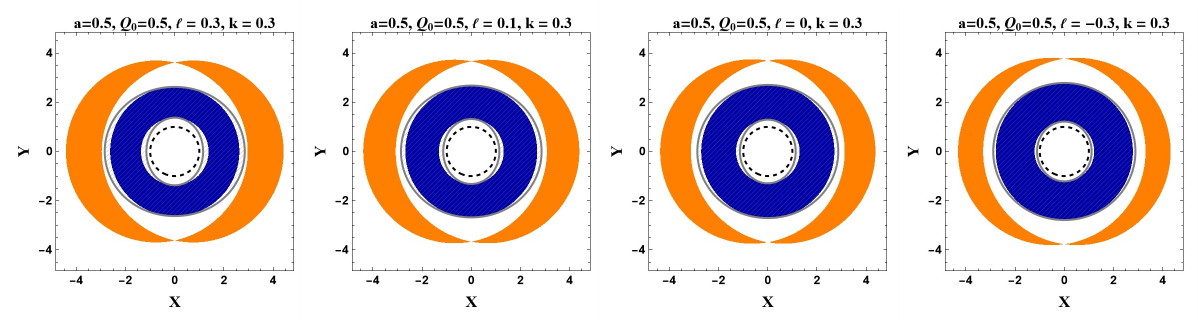}
\caption{The photon regions of the KN-like black hole in Bumblebee gravity surrounded by plasma in the $(r, \theta)$ plane. The parameters are set to $a = 0.5$, $Q_0 = 0.5$, and $k = 0.3$, with varying Lorentz-violation parameter $\ell$.}
\label{fig4a}
\end{figure*}

Substituting the plasma model into eq.\eqref{3.7}, we can obtain explicit expressions for the radial and angular effective potentials, $R(r)$ and $\Theta(\theta)$:
\begin{equation}\label{3.10}
\begin{aligned}
\frac{\Delta_r}{1+\ell} \rho^2 \dot{r}^2 &= \frac{\Delta_r}{1+\ell} R(r) \\
&= \frac{[(r^2+a_{\text{eff}}^2)E - a_{\text{eff}}L_z]^2}{\Delta_r} \\
&\quad - (L_z - a_{\text{eff}}E)^2 - \mathcal{K} - \omega_c^2 \sqrt{r}, \\[1ex]
\rho^2 \dot{\theta}^2 &= \Theta(\theta) \\
&= \mathcal{K} + \left[a_{\text{eff}}^2 E^2 - \frac{L_z^2}{\sin^2\theta}\right] \cos^2\theta.
\end{aligned}
\end{equation}

To establish the connection between the photon's orbit and the shadow seen by a distant observer, it is necessary to write the explicit geodesic equation for the photon. From Hamilton's canonical equations
\begin{equation}\label{3.11}
\dot{x}^\mu = \frac{\partial H}{\partial p_\mu}, \quad \dot{p}_\mu = -\frac{\partial H}{\partial x^\mu}.
\end{equation}

The first-order geodesic equations for photons are obtained as follows:
\begin{equation}\label{3.12}
\begin{aligned}
\rho^2 \dot{t} &= \frac{a_{\text{eff}} L_z [\Delta_r - (r^2 + a_{\text{eff}}^2)] + (r^2 + a_{\text{eff}}^2)^2 E}{\Delta_r} \\
&\quad - a_{\text{eff}}^2 E \sin^2 \theta, \\[1ex]
\rho^2 \dot{r} &= \sqrt{R(r)}, \\[1ex]
\rho^2 \dot{\theta} &= \sqrt{\Theta(\theta)}, \\[1ex]
\rho^2 \dot{\phi} &= \frac{a_{\text{eff}} E [(r^2 + a_{\text{eff}}^2) - \Delta_r]}{\Delta_r} \\
&\quad + \frac{L_z (\Delta_r - a_{\text{eff}}^2 \sin^2 \theta)}{\Delta_r \sin^2 \theta}.
\end{aligned}
\end{equation}

\begin{figure*}[t]
\centering
\includegraphics[width=0.40\textwidth]{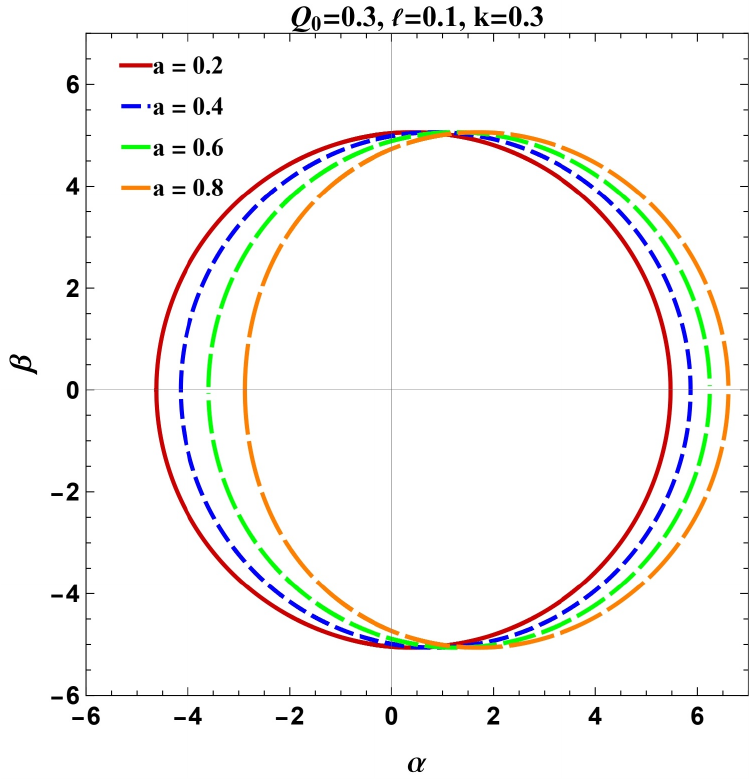}
\includegraphics[width=0.40\textwidth]{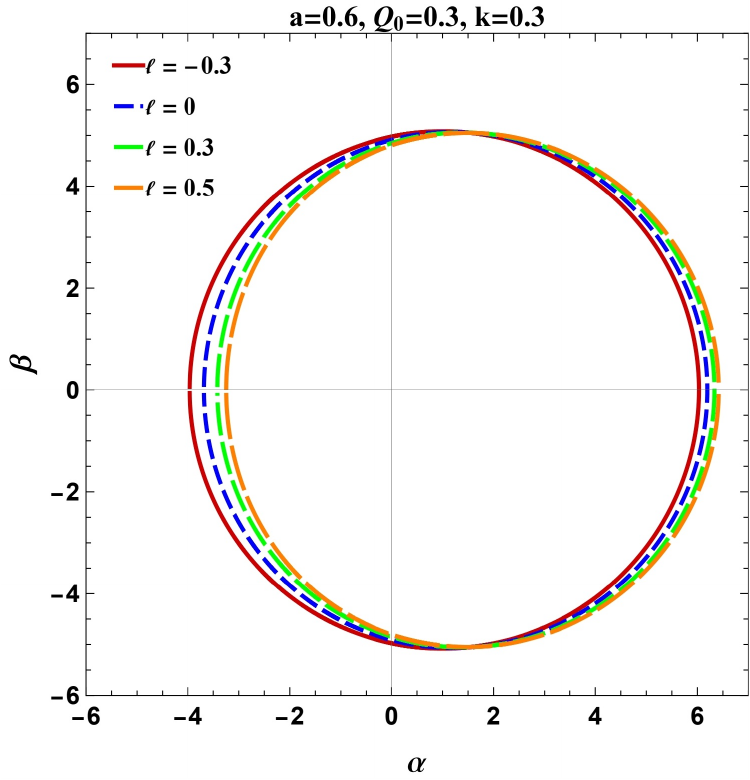}

\includegraphics[width=0.40\textwidth]{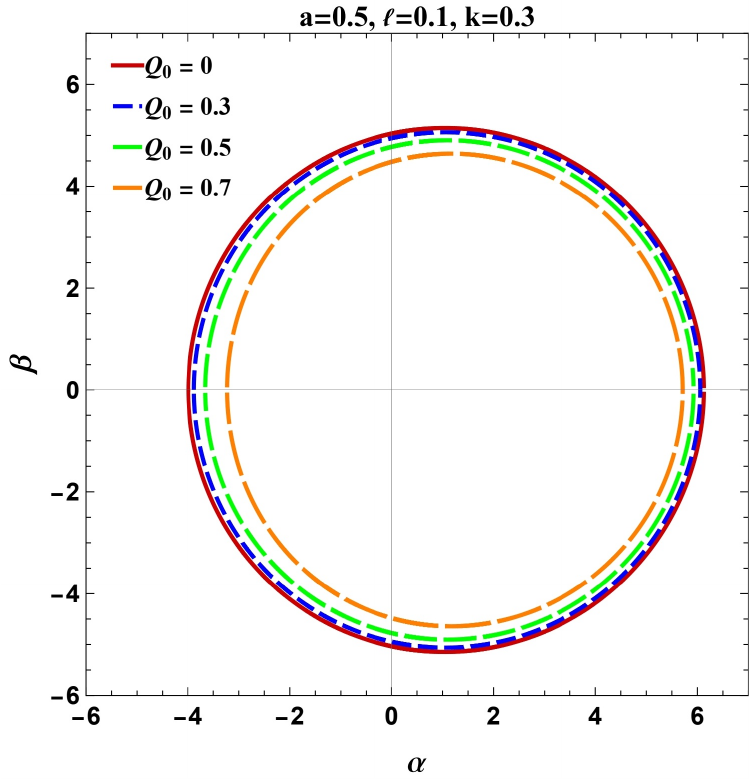}
\includegraphics[width=0.40\textwidth]{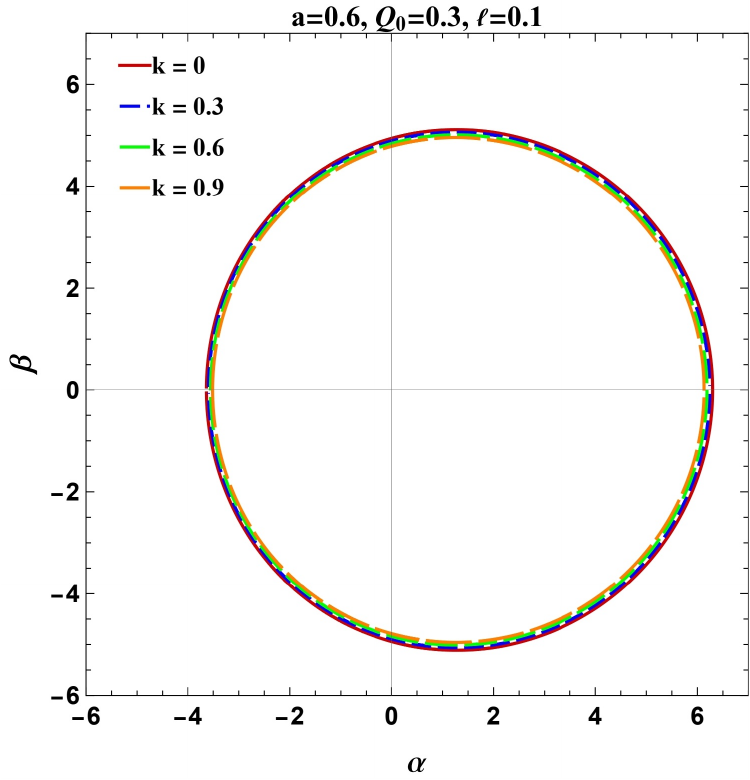}
\caption{Shadows of the KN-like black hole in Bumblebee gravity surrounded by plasma, illustrating the independent effects of varying the parameters $a$, $\ell$, $Q_0$, and $k$.}
\label{fig5}
\end{figure*}

\begin{figure*}[t]
\centering
\includegraphics[width=0.5\textwidth]{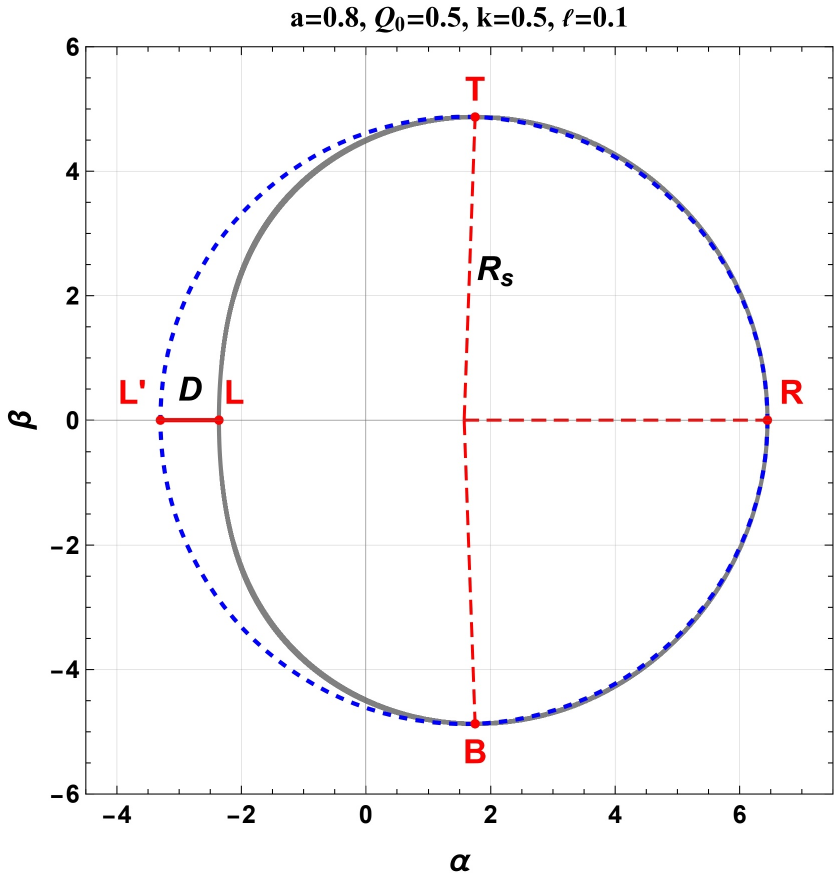}
\caption{Illustration of shadow observables defined by the reference circle method. The solid black curve represents the black hole shadow, and the dashed blue curve denotes the reference circle. This circle is constructed to pass through three characteristic points: the top ($T$), bottom ($B$), and rightmost ($R$) points of the shadow, while $L$ and $L'$ denote the left endpoints of the shadow and the reference circle, respectively. The parameter $D$ is defined as $D\equiv \alpha_{L'}-\alpha_{L}$.}
\label{fig6}
\end{figure*}

Next, two dimensionless impact parameters are introduced:
\begin{equation}\label{3.16}
\xi = \frac{L_z}{E}, \quad \eta = \frac{K}{E^2},
\end{equation}
which correspond to the angular momentum and the generalized carter constant normalized by the energy, respectively.

The boundary of the black hole shadow is determined by the unstable spherical photon orbits. Their radial motion satisfies the following critical conditions:
\begin{equation}\label{3.17}
R(r) = 0 , \quad R(r)' = 0
\end{equation}
and
\begin{equation}\label{3.18}
\Theta(\theta) \geq 0 \quad \text{as} \quad \theta \in [0, \pi].
\end{equation}

Solving these conditions simultaneously, we obtain the explicit expressions for the impact parameters $\xi$ and $\eta$:
\begin{align}
\xi &= \frac{r^2 + a_{\text{eff}}^2}{a_{\text{eff}}} - \frac{\Delta_r}{a_{\text{eff}} \Delta'_r} \left( 2r + \sqrt{4r^2 - \frac{1}{2} \Delta'_r k r^{-1/2}} \right) \label{3.19} \\[1ex]
\eta &= \frac{\Delta_r}{(\Delta'_r)^2} \left( 2r + \sqrt{4r^2 - \frac{1}{2} \Delta'_r k r^{-1/2}} \right)^2 \notag \\
&\quad - \frac{1}{a_{\text{eff}}^2} \left[ r^2 - \frac{\Delta_r}{\Delta'_r} \left( 2r + \sqrt{4r^2 - \frac{1}{2} \Delta'_r k r^{-1/2}} \right) \right]^2 \notag \\
&\quad - k \sqrt{r}. \label{3.20}
\end{align}
Substituting eqs.\eqref{3.19} and \eqref{3.20} into eq.\eqref{3.18}, we obtain the existence condition for the photon region:
\begin{equation}\label{3.21}
(\Delta_r X - \rho^2 \Delta'_r)^2 \leq a_{\text{eff}}^2 \sin^2 \theta \left[ \Delta_r X^2 - k\sqrt{r} (\Delta'_r)^2 \right],
\end{equation}
where
\begin{equation}\label{3.22}
X = 2r + \sqrt{4r^2 - \frac{1}{2} \Delta'_r k r^{-1/2}}.
\end{equation}

Unstable spherical null geodesics correspond to the boundary of the black hole shadow. Therefore, one must analyze the stability of photon motion against radial perturbations at $r = r_p$. This stability is determined by the sign of the second derivative of the effective potential $R(r)$. Utilizing previously derived eqs.\eqref{3.10} and \eqref{3.19}, \eqref{3.20}, we obtain
\begin{equation}\label{3.23}
\begin{split}
\frac{\mathcal{R}''(r)}{E^2} (\Delta'_r)^2 &= 4\Delta_r \Delta'_r X + 8r^2 (\Delta'_r)^2 \\
&\quad - \Delta_r \Delta''_r X^2 - \frac{k(\Delta'_r)^2}{4r^{3/2}} (4r\Delta'_r - \Delta_r),
\end{split}
\end{equation}

For the vacuum case $k=0$, $X =4r$, the formula reduces to
\begin{equation}
\frac{R''(r)}{8 E^2} (\Delta'_r)^2 = 2r \Delta_r \Delta'_r + r^2 (\Delta'_r)^2 - 2r^2 \Delta_r \Delta''_r.
\end{equation}

The condition $R''(r_p) > 0$ indicates an unstable spherical null geodesic, whereas ${R''(r_p) < 0}$ implies a stable one.

The photon regions of the KN-like black hole in Bumblebee gravity surrounded by plasma are plotted in the $(r, \theta)$ plane, as depicted in figures~\ref{fig3a} and~\ref{fig4a}, where orange region represents unstable photon orbits and yellow region represents stable photon orbits. According to refs. \cite{Sun:2024xtf,Meng:2022kjs,Grenzebach:2014fha}, we adopt two different scales along the $r$-direction to depict the entire spacetime, emphasizing the exterior areas $r > 0$. Specifically, the radial coordinate is scaled as $r + M$ in the region $r > 0$ and $M \exp(r/M)$ in the region $r < 0$. In this mapped space, the black hole throat at $r = 0$ is indicated by a black dashed circle, while the blue ($\Delta_r \leq 0$) and gray ($g_{tt} > 0$) regions correspond to the event horizon and the ergosphere, respectively.

In figure~\ref{fig3a}, we fix $a=0.8$, $Q_0=0.5$, and $\ell=0.1$ to observe the impact of the plasma parameter $k$ on the photon regions. We can clearly see an exterior photon region outside the outer horizon. Meanwhile, there sometimes exist unstable and stable photon orbits in the interior photon region inside the inner horizon, which are symmetric with respect to the equatorial plane. As $k$ increases, the exterior photon region remains almost unchanged, but the interior photon region gradually diminishes until it vanishes completely. Figure~\ref{fig4a} illustrates the influence of the Lorentz-violating parameter $\ell$ on the photon regions, with the background parameters fixed at $a=0.5$, $Q_0=0.5$, $k=0.3$. As $\ell$ decreases, the exterior unstable photon region (orange area) shrinks. Furthermore, the blue region representing the event horizon expands, while the gray region representing the ergosphere closely clings to the exterior of the horizon.

For an observer positioned at infinity $r_0 \to \infty$ with an inclination angle $\theta_0$, we introduce the celestial coordinates $(\alpha, \beta)$. The mapping relation between these coordinates and the impact parameters $(\xi, \eta)$ is given by
\begin{equation}\label{3.30}
\alpha = -\frac{\xi}{\sin \theta_0},
\end{equation}
\begin{equation}
\beta = \pm \sqrt{\eta + a_{\text{eff}}^2 \cos^2 \theta_0 - \xi^2 \cot^2 \theta_0}.
\end{equation}

Consequently, parametrized by the radius $r_p$ of the unstable spherical photon orbits, this mapping defines the locus of a closed curve on the observer's sky, which constitutes the shadow of the Bumblebee black hole surrounded by plasma. Before proceeding with the shadow calculation, we first check whether the radial function $\Delta(r) = 0$ possesses real positive roots for each set of model parameters. This step ensures that the corresponding spacetime contains an event horizon. Parameter points that fail to satisfy the condition for the existence of an event horizon are directly excluded during the calculation process.

\section{The black hole shadow and shadow observables}
\label{The black hole shadow and shadow observables}

\begin{figure*}[t]
\centering
\includegraphics[width=0.45\textwidth]{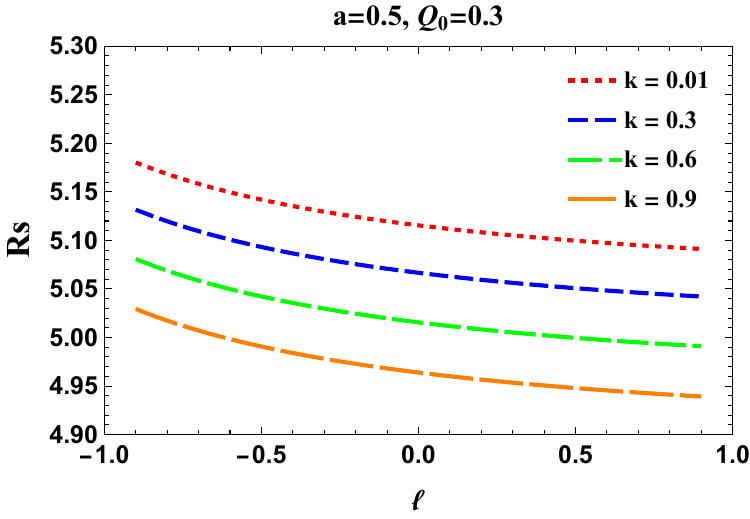}
\includegraphics[width=0.45\textwidth]{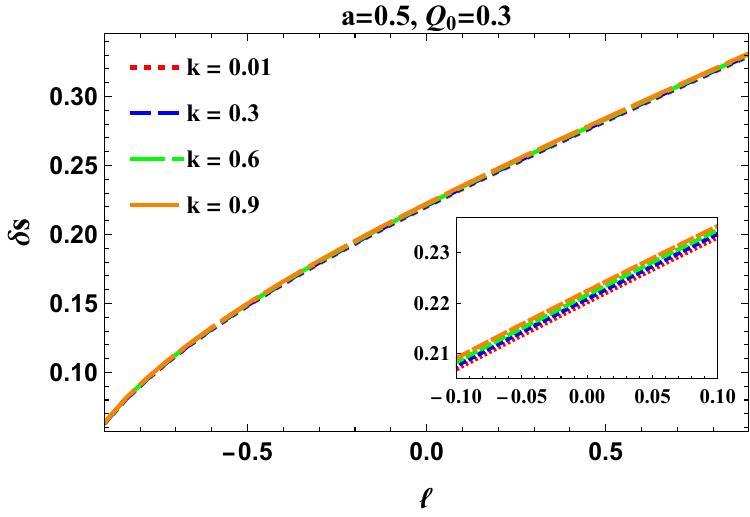}

\includegraphics[width=0.45\textwidth]{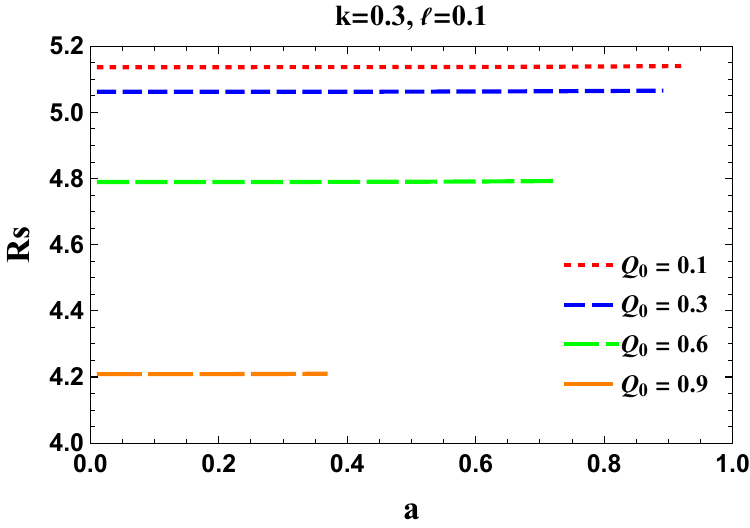}
\includegraphics[width=0.45\textwidth]{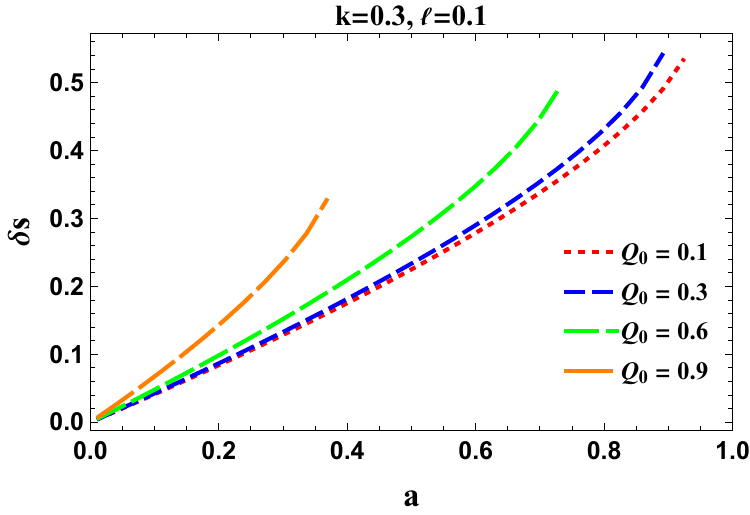}
\caption{Variations of the shadow radius $R_s$ and the distortion parameter $\delta_s$ for a KN-like black hole in Bumblebee gravity surrounded by plasma. The fixed parameters are set to $a=0.5$, $Q_0=0.1$ (the upper row) and \mbox{$k=0.3$, $\ell=0.1$} (the bottom row).}
\label{fig7}
\end{figure*}

\begin{figure*}[t]
\centering
\includegraphics[width=0.45\textwidth]{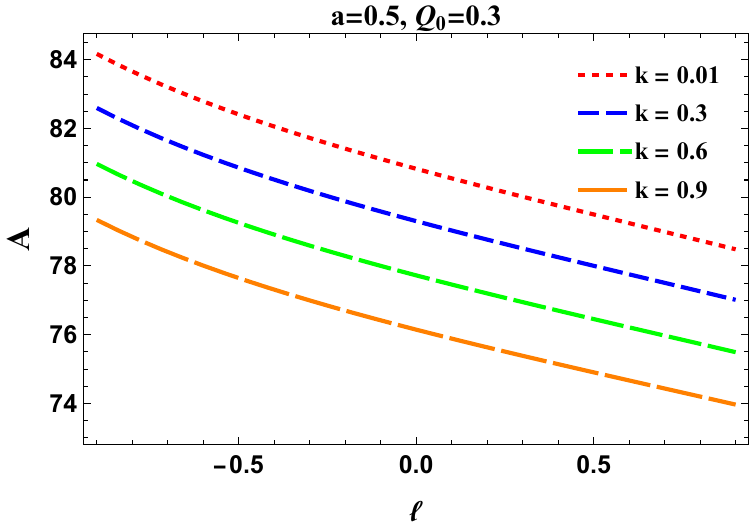}
\includegraphics[width=0.45\textwidth]{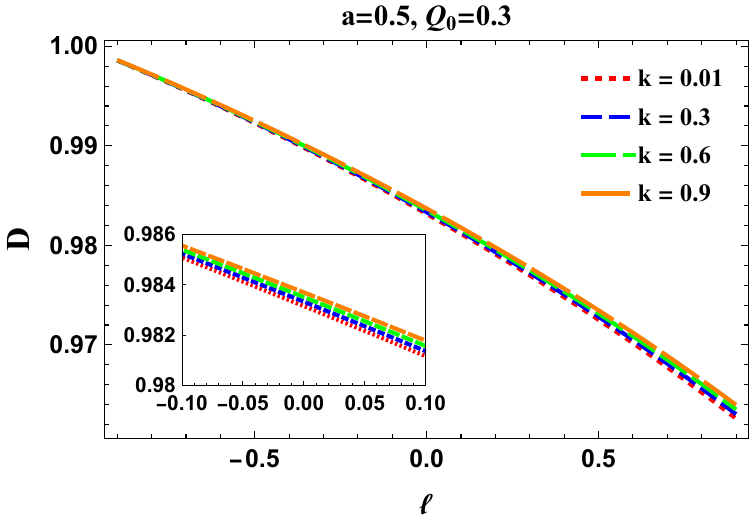}

\includegraphics[width=0.45\textwidth]{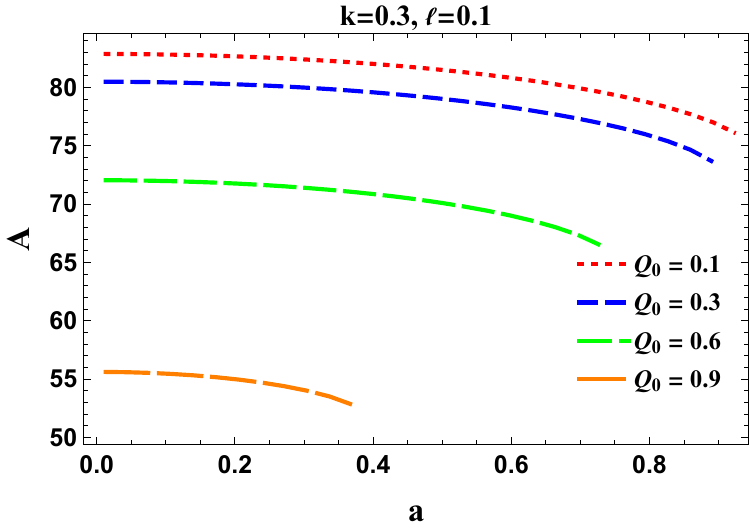}
\includegraphics[width=0.45\textwidth]{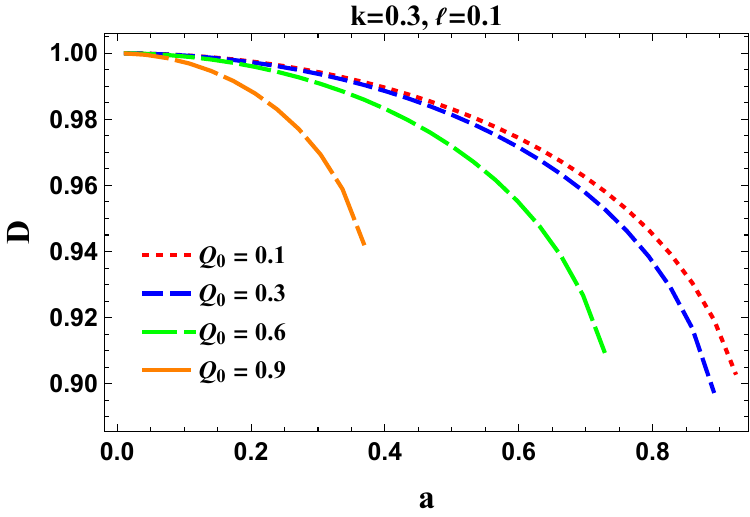}
\caption{Variations of the shadow area $A$ and the oblateness $D$ for a KN-like black hole in Bumblebee gravity surrounded by plasma. The fixed parameters are set to $a=0.5$, $Q_0=0.1$ (the upper row) and \mbox{$k=0.3$, $\ell=0.1$} (the bottom row).}
\label{fig8}
\end{figure*}

The shadow of the KN-like black hole in Bumblebee gravity surrounded by plasma is governed by the combined influence of the plasma parameter $k$ and the Lorentz-violation parameter $\ell$. In this section, we proceed to analyze the independent impact of each parameter on both the shadow and its observables. Subsequently, a brief analysis of the energy emission rate is performed.

To understand the independent influence of each physical parameter on the shadow, we examine the response to variations in a single parameter while keeping the others fixed, as illustrated in figure~\ref{fig5}. The results indicate that the effects can be broadly classified into two categories: On the one hand, increasing the spin parameter $a$ and the Lorentz-violation parameter $\ell$ primarily enhances the left-right asymmetry of the shadow relative to the $\alpha=0$ axis, leading to more pronounced distortion. On the other hand, increasing the charge parameter $Q_{0}$ or the plasma parameter $k$ predominantly leads to a reduction in the shadow size.

Since these parameters significantly alter the shadow's shape and size, to effectively characterize the shadow, it is imperative to introduce a set of observables that can precisely describe the black hole shadow geometry.

Following the reference circle method proposed by Hioki and Maeda \cite{Hioki:2009na}, we quantify the shadow's size and shape  as illustrated in figure~\ref{fig6}. By constructing a reference circle passing through three characteristic points—the top ($T$), bottom ($B$), and rightmost ($R$) points of the shadow boundary, we can define the equivalent shadow radius and the distortion parameter as follows \cite{Hioki:2009na}:

\begin{equation}
R_{s}=\frac{(\alpha_{t}-\alpha_{r})^{2}+\beta_{t}^{2}}{2\left|\alpha_{t}-\alpha_{r}\right|},
\quad
\delta_{s}=\frac{D_{cs}}{R_{s}}=\frac{\left|\alpha_{L}-\alpha_{L}^{\prime}\right|}{R_{s}}.
\label{eq:Rs-deltas}
\end{equation}

Figure~\ref{fig7} presents the behavioral profiles of the shadow radius $R_s$ and the distortion parameter $\delta_s$ under the variations of different physical parameters. The shadow radius $R_s$ is primarily reduced by increasing the plasma parameter $k$ and charge parameter $Q_0$. It exhibits a slight decrease with the Lorentz-violating parameter $\ell$, but remains almost unchanged with the spin parameter $a$. In contrast, the distortion parameter $\delta_s$ is significantly enhanced with increasing in $\ell$ and $a$. Furthermore, while $\delta_s$ is practically insensitive to $k$, it experiences a slight enhancement at larger values of $Q_0$.

Since $R_s$ and $\delta_s$ require specific symmetries, they are not well-suited for highly distorted black hole shadows. To characterize shadow geometries with more general shapes, we further introduce two additional observables proposed by Kumar and Ghosh \cite{Rahul:2020pcy}, the shadow area $A$ and the oblateness $D$, which are defined as follows \cite{Rahul:2020pcy}:
\begin{equation}
\begin{split}
A &= 2 \int \beta(r_p) d\alpha(r_p) = 2 \int_{r_p^{min}}^{r_p^{max}} \left( \beta(r_p) \frac{d\alpha(r_p)}{dr_p} \right) dr_p, \\[2ex]
D &= \frac{\alpha_r - \alpha_l}{\beta_t - \beta_b}.
\end{split}
\end{equation}

Here, $r_p^{min}$ and $r_p^{max}$ represent the radii of the prograde and retrograde photon orbits, determined by $\eta(r_p) = 0$. For an observer on the equatorial plane, the oblateness $D$ \cite{Tsupko:2017rdo} varies within the interval $[\sqrt{3}/2, 1]$. Specifically, the upper limit $D=1$ denotes the spherically symmetric Schwarzschild limit, whereas the lower bound $D=\sqrt{3}/2$ characterizes the extremal Kerr black hole.

Figure~\ref{fig8} illustrates the behavioral profiles of the shadow area $A$ and the oblateness $D$ under the variations of different physical parameters. As depicted in the upper panels, an increase in the Lorentz-violating parameter $\ell$ leads to a reduction in both the shadow area $A$ and the oblateness $D$. Moreover, as the plasma parameter $k$ increases, the shadow area shrinks significantly, whereas the oblateness experiences a slight increase. Conversely, the lower panels reveal the influence of the spin parameter $a$ and charge parameter $Q_0$. With a growing spin parameter $a$, the oblateness $D$ drops sharply, and the shadow area $A$ has a slight reduction. Furthermore, increasing the charge parameter $Q_0$ leads to a significant decrease in the shadow area and a reduction in the oblateness.

In addition to the geometric shadow properties, the high-frequency energy emission rate of a black hole in the geometric optics limit can also reflect its spacetime structure. As a supplement to the shadow analysis above, we further investigate the high-frequency emission rate of the KN-like black hole in a plasma background. For an observer at infinity, the high-frequency energy emission rate of the black hole can be expressed as:
\begin{equation}
\frac{d^{2}E(\omega)}{d\omega\, dt}
=\frac{2\pi^{2}\sigma_{\rm lim}}{e^{\omega/T_{H}}-1}\,\omega^{3}.
\end{equation}

Here, $\omega$ is the photon frequency and $T_{H}$ is the Hawking temperature. $\sigma_{\rm lim}$ stands for the high-frequency absorption cross-section. In the geometric optics limit, its oscillation tends to a constant value, which corresponds to the geometric shadow area of the black hole \cite{Wei:2013kza}:
\begin{equation}
\sigma_{\rm lim}\approx \pi R_{s}^{2},
\end{equation}
where $R_{s}$ is the black hole shadow radius obtained above.

\begin{figure*}[t]
\centering
\includegraphics[width=0.45\textwidth]{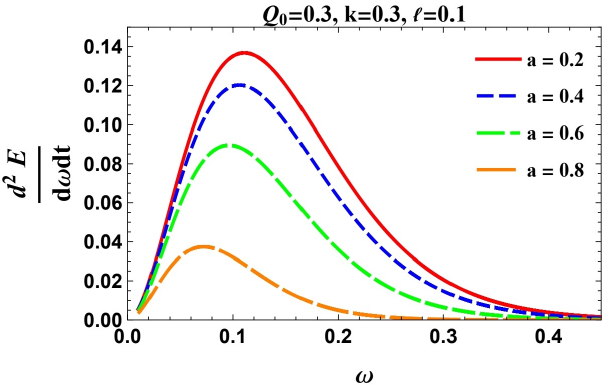}
\includegraphics[width=0.45\textwidth]{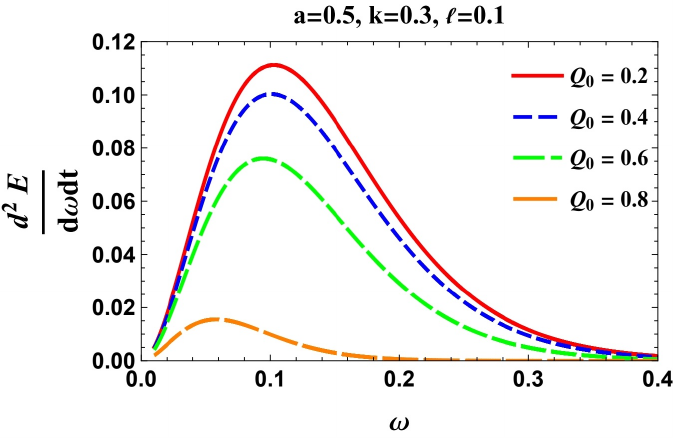}

 \includegraphics[width=0.45\textwidth]{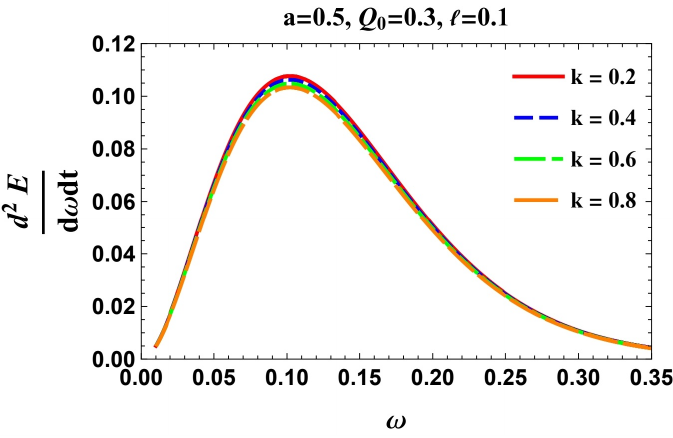}
\includegraphics[width=0.45\textwidth]{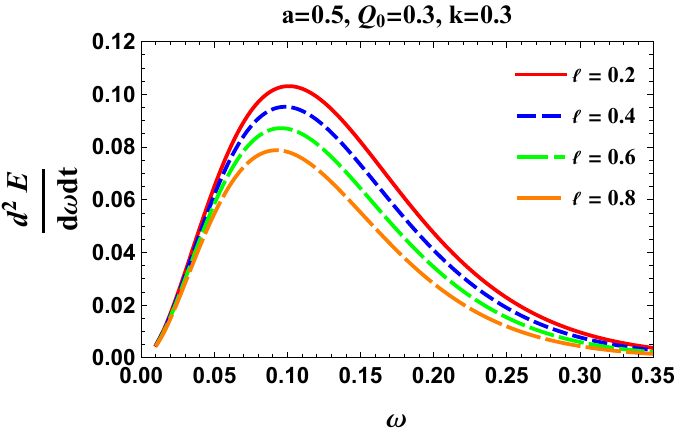}
\caption{High-frequency energy emission rate as a function of frequency $\omega$ for varying physical parameters ($a, Q_0,k,\ell$).}
\label{fig9}
\end{figure*}

The Hawking temperature $T_{H}$ in the KN-like black hole background is given by:
\begin{equation}
T_{H}=\frac{r_{+}-M}{2\pi\left(r_{+}^{2}+a^{2}(1+\ell)\right)}.
\end{equation}

 Figure~\ref{fig9} displays the response of the energy emission rate to various parameters while keeping the others fixed. The upper row plots show that the peak of the energy emission rate significantly decreases as $a$ or $Q_0$ increases, and the peak shifts to lower frequency region. While the bottom row reveals that although an increase in $k$ or $\ell$ also suppresses the energy emission rate, this reduction is relatively mild.

\section{Constraints from EHT Observations on M87*}
\begin{figure*}[t]
\centering
\includegraphics[height=0.291\textwidth]{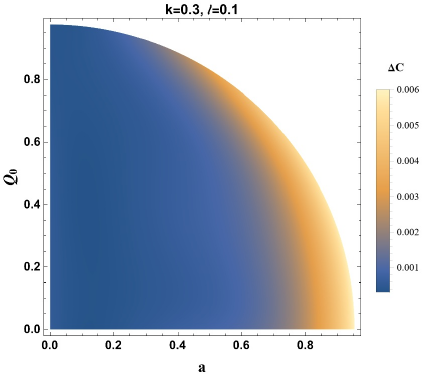} \hfill
\includegraphics[height=0.291\textwidth]{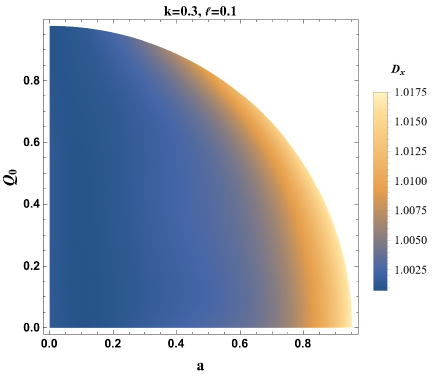} \hfill
\includegraphics[height=0.291\textwidth]{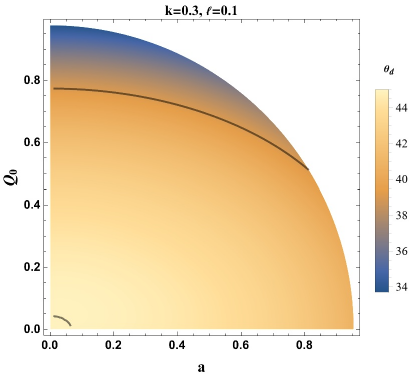}
\caption{Density plots of the circular deviation $\Delta C$, axial ratio $D_x$, and angular diameter $\theta_d$ in the ($a, Q_0$) parameter plane for $\theta_0 = 17^\circ$.The parameter plane contains only parameter values for which an event horizon exists.}
\label{fig10}
\end{figure*}

\vspace{0.5cm} %

\begin{figure*}[t]
\centering
\includegraphics[height=0.2875\textwidth]{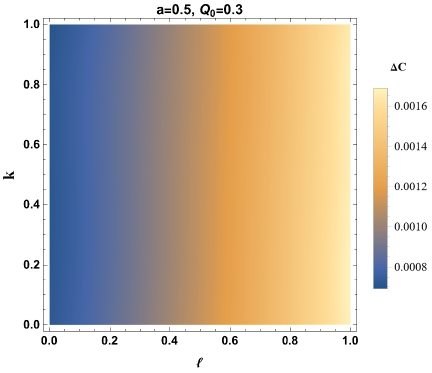} \hfill
\includegraphics[height=0.2875\textwidth]{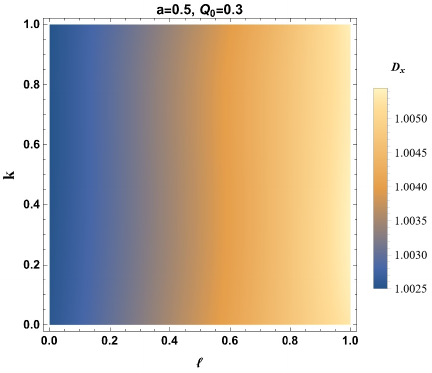} \hfill
\includegraphics[height=0.2875\textwidth]{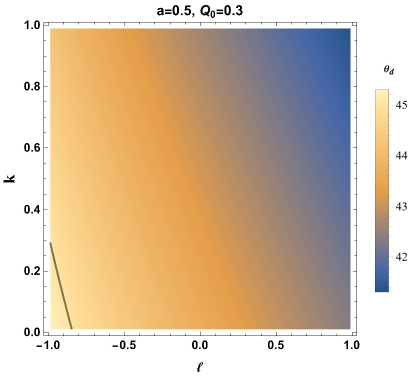}
\caption{Density plots of the circular deviation $\Delta C$, axial ratio $D_x$, and angular diameter $\theta_d$ in the ($\ell, k$) parameter plane for $\theta_0 = 17^\circ$.The parameter plane contains only parameter values for which an event horizon exists.}
\label{fig11}
\end{figure*}
The shadow image of the supermassive black hole M87* captured by the EHT collaboration reveals a unique crescent-shaped emission ring structure. More importantly, based on the estimated inclination angle of the M87* observation $\theta_0 = 17^\circ$, the EHT collaboration extracted key geometric constraints from this image \cite{EventHorizonTelescope:2019dse,EventHorizonTelescope:2019ths,EventHorizonTelescope:2019pgp}, including the circularity deviation $\Delta C \lesssim 0.1$, axis ratio $1 < D_x \lesssim \frac{4}{3}$, and angular diameter $\theta_d = 42 \pm 3 \, \mu\text{as}$.

\begin{figure*}[t]
\centering
\includegraphics[width=0.45\textwidth]{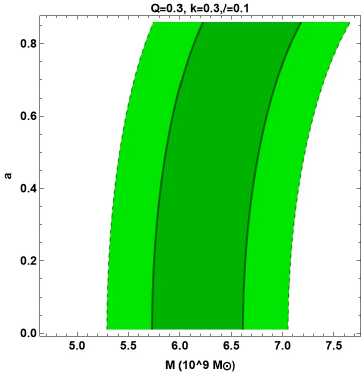}
\includegraphics[width=0.45\textwidth]{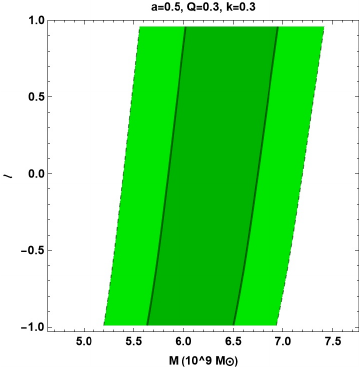}
\caption{Constraints on parameter $a$, $\ell$ and estimated M87* black hole mass $M (\times 10^9 M_{\odot})$ using M87* shadow angular size within $1\sigma$ (dark green region) and $2\sigma$ (light green region).}
\label{fig12}
\end{figure*}

In this section, we consider the M87* black hole as a Kerr-Newman-like (KN-like) black hole surrounded by plasma within the framework of Bumblebee gravity, and use the aforementioned EHT observational constraints to restrict the model parameters, thereby identifying the physically feasible regions of the model.

In the celestial observer's plane, the shadow can be represented by polar coordinates $(R(\varphi), \varphi)$, with its origin at the center of the shadow $(\alpha_C, \beta_C)$. Next, we give the expressions for $\Delta C$, $D_x$, and $\theta_d$. First, define the shadow center point:
\begin{equation}
\beta_C = 0, \quad \alpha_C = \frac{\alpha_R + \alpha_L}{2},
\end{equation}
where $\alpha_R$ and $\alpha_L$ are the $\alpha$ coordinates of the right and left edges of the shadow, respectively. Next, represent the shadow boundary in polar coordinates:
\begin{equation} \label{3.33}
\begin{split}
R(\varphi) &= \sqrt{(\alpha - \alpha_C)^2 + (\beta - \beta_C)^2}, \\[1ex]
\varphi &= \tan^{-1} \left( \frac{\beta}{\alpha - \alpha_C} \right).
\end{split}
\end{equation}

Define the average radius \cite{Bambi:2019tjh}:
\begin{equation}
\bar{R}^2 = \frac{1}{2\pi} \int_0^{2\pi} R^2(\varphi) \, d\varphi.
\end{equation}

The circularity deviation is then defined as \cite{Afrin:2021imp}
\begin{equation}
\Delta C = \frac{1}{\bar{R}} \sqrt{\frac{1}{2\pi} \int_0^{2\pi} (R(\varphi) - \bar{R})^2 \, d\varphi}.
\end{equation}

The axial ratio is defined as the vertical diameter of the shadow divided by the horizontal diameter \cite{Banerjee:2019nnj}
:
\begin{equation}
D_x = \frac{\beta_t - \beta_b}{\alpha_r - \alpha_l}.
\end{equation}

The formula for the angular diameter is as follows \cite{Kumar:2020owy}:
\begin{equation}
\theta_d = \frac{2R_a}{d}, \quad R_a = \sqrt{\frac{A}{\pi}},
\end{equation}
where $d$ is the distance of M87* from Earth, $R_a$ can be called the shadow areal radius, and $A$ is the shadow area.

Next, we model the supermassive black hole M87* as a KN-like black hole within the framework of Bumblebee gravity, immersed in a non-homogeneous power-law plasma. By setting the mass $M = 6.5 \times 10^9 M_\odot$ and the distance $d = 16.8 \, \text{Mpc}$, we can determine the theoretical observables for our model. Furthermore, we adopt an inclination angle of $\theta_0 = 17^\circ$ for M87*, which is estimated based on the orientation of the relativistic jet \cite{CraigWalker:2018vam}.

\label{Constraints from EHT Observations on M87*}

Figure~\ref{fig10} shows the comparison between the theoretical shadow predictions of our black hole model and the EHT observational constraints for M87* with ${\theta_0 = 17^\circ}$ in the ($a, Q_0$) parameter plane. The two solid black curves represent the constant angular diameters of ${\theta_d = 39  \, \mu\text{as}}$ and ${\theta_d = 45  \, \mu\text{as}}$, respectively. The region between them denotes the parameter space consistent with the EHT constraints. More specifically, when $k=0.3$ and $\ell=0.1$, the EHT observation ${\theta_d = 42 \pm 3 \, \mu\text{as}}$ rules out part of the parameter space excluding the regions corresponding to a minimum spin-charge combination of ($a < 0.06$, $Q_0 < 0.04$) and a maximum charge constraint of $Q_0 > 0.77$. By contrast, the EHT observational constraints on the circularity deviation $\Delta C \lesssim 0.1$ and axis ratio $1 < D_x \lesssim 4/3$ are satisfied in all the considered parameter regions, which implies that it is difficult to use circularity deviation and axis ratioin to constrain the parameters of the KN-like black hole. 

The density plots in the ($\ell, k$) parameters plane are presented in figure~\ref{fig11} when $a=0.5$ and $Q_0=0.3$. On the one hand, all parameter regions satisfy the constraints $\Delta C \lesssim 0.1$ and  $1 < D_x \lesssim 4/3$, indicating consistency with the EHT observations. On the other hand, the solid black curve corresponds to the angular diameter ${\theta_d = 45  \, \mu\text{as}}$ in the density plot. The region with $0 < k < 0.3$ and $-1 < \ell < -0.85$ is therefore excluded by the EHT angular diameter constraint.

Therefore, we conclude that EHT observations of the M87* black hole shadow cannot rule out the existence of a KN-like black hole in Bumblebee gravity surrounded by plasma. To further constrain the spin parameter $a$ and the Lorentz-violation parameter $\ell$, the $1\sigma$ and $2\sigma$ confidence intervals are plotted based on the aforementioned angular diameter density maps. Figure~\ref{fig12} shows that the adopted mass $M = 6.5 \times 10^9 M_{\odot}$ falls within the $1\sigma$ band, indicating that our setup is consistent with the EHT constraints.
For $Q_0=0.3, k=0.3, \ell=0.1$, the $1\sigma$ bound on mass is \mbox{$5.72 \times 10^{9} M_{\odot} < M < 7.18 \times 10^{9} M_{\odot}$} and the $2\sigma$ bound is \mbox{$5.29 \times 10^{9} M_{\odot} < M < 7.66 \times 10^{9} M_{\odot}$}, while for $a=0.5, Q_0=0.3, k=0.3$, the $1\sigma$ bound on mass is \mbox{$5.62 \times 10^{9} M_{\odot} < M < 6.94 \times 10^{9} M_{\odot}$} and the $2\sigma$ bound is \mbox{$5.20 \times 10^{9} M_{\odot} < M < 7.41 \times 10^{9} M_{\odot}$}.

\section{Conclusion}
\label{Conclusion}
To summarize, in this work, we investigated the shadow of a charged rotating black hole within the framework of Bumblebee gravity, immersed in a non-homogeneous power-law plasma model. In order to analyze the shadow characteristics under the combined effect of Lorentz symmetry breaking (LSB) and the plasma medium, we studied the black hole shadow and its associated observables. Furthermore, we utilized the latest EHT observations to constrain the model parameters.

First, we explored the influence of the Lorentz-violating parameter $l$ and the charge parameter $Q_0$ on the event horizon through an analysis of the parameter space and spacetime structure. As shown in figure~\ref{fig1}, with the increase of $l$ and $Q_0$, the allowable range of the spin parameter $a$ is significantly compressed, and the spacetime structure exhibits a strong tendency to approach an extremal black hole. This fundamental geometric analysis not only established the physical boundaries for subsequent simulations but also indicates that the LSB provides a non-trivial correction to the properties of the strong gravitational field.

Building upon this, we derived the null geodesic equations for photons in a plasma model using the separable Hamilton-Jacobi equation. Along with the analysis of the photon regions, we numerically calculated the black hole shadow. By systematically comparing the effects of individual parameters on the shadow and its observables ($R_s, \delta_s$), we categorized the roles of the four physical parameters into two distinct physical effects: spin $a$ and the Lorentz-violating parameter $l$ mainly cause a left-right asymmetry in the shadow relative to the $\alpha=0$ axis, dominating the drag distortion of the shadow, while charge parameter $Q_0$ and the plasma parameter $k$ dominate the radial contraction of the shadow size.

Finally, we mapped three key observables of the M87* black hole—the angular diameter, axial ratio, and circularity deviation—onto our theoretical model. Figures~\ref{fig10} and \ref{fig11} demonstrated that, although the current observational precision is insufficient to completely rule out the validity of the Bumblebee gravity model,  we successfully excluded a portion of the parameter space characterized by a low Lorentz-violating parameter and low plasma density. As shown in figure~\ref{fig12}, the confidence intervals analysis demonstrated that our proposed model is self-consistent under the EHT constraints. This result not only verified the feasibility of using black hole images to test fundamental physics but also provided a theoretical reference for future, higher-precision gravitational tests.

\section*{Acknowledgments}

The work was supported by Yunnan Xingdian Talent Support Program-Young Talent Project.

\end{document}